\documentclass[11pt]{article}
\usepackage[margin=1in]{geometry}
\usepackage{amsmath,amssymb,amsthm,mathtools}
\usepackage[T1]{fontenc}\usepackage{mathptmx}\usepackage[expansion=false]{microtype}
\usepackage{graphicx,booktabs,enumitem,setspace}
\usepackage[numbers,sort&compress]{natbib}
\usepackage[hidelinks]{hyperref}
\newcommand{\E}{\mathbb{E}}
\newcommand{\cov}{\operatorname{cov}}
\newcommand{\var}{\operatorname{var}}
\newcommand{\dd}{\,\mathrm{d}}
\newcommand{\ind}{\mathbf{1}}

\newcommand{\indep}{\perp\!\!\!\perp}
\newcommand{\pr}{\mathrm{pr}}
\newcommand{\IV}{\mathrm{IV}}
\newcommand{\ACE}{\mathrm{ACE}}

\theoremstyle{plain}
\newtheorem{result}{Result}
\newtheorem{proposition}{Proposition}
\theoremstyle{definition}
\newtheorem{condition}{Condition}

\title{\bfseries What does two-sample Mendelian randomization estimate when the exposure--outcome relationship is nonlinear?}
\author{Eric J. Tchetgen Tchetgen\\[2pt]
\small Department of Statistics and Data Science, The Wharton School, University of Pennsylvania\\
\small \texttt{ett@wharton.upenn.edu}}
\date{\today}

\begin{document}
\maketitle
\onehalfspacing

\begin{abstract}
\noindent Two-sample Mendelian randomization with summary statistics has become a popular MR design in Epidemiology. Its ratio estimate has an average causal effect interpretation only under strong untestable assumptions including linearity. We derive what the per-variant ratio, and the pooled and robust estimators built from it, estimate when a continuous exposure has a nonlinear effect, under two conditions on how genetic variation and unmeasured factors shape exposure: no unmeasured common effect modifier, and an exposure control function. Under either condition, each variant's ratio is a weighted average of the slope of the dose-response curve, with weights that depend on how that variant moves the distribution of exposure and that are nonnegative when the exposure distributions at different genotypes do not cross. Variants that shift everyone's exposure equally target nearly the same quantity; variants acting at particular doses or on the spread of exposure target different ones, some outside the range of any individual effect. Pooled estimates therefore have no fixed interpretation, heterogeneity tests and MR-Egger misread nonlinearity as pleiotropy, and summary statistics cannot reveal the problem, even when the dose-response curve is identifiable from individual-level data. For continuous exposures, two-sample estimates rest on a linearity assumption that should be stated and defended; individual-level exposure data resolve the problem and should be prioritized.
\end{abstract}


\section{Introduction}

Mendelian randomization uses genetic variants as instruments for a modifiable exposure~\cite{DaveySmith2003,DaveySmithHemani2014}. The two-sample design~\cite{Pierce2013,Burgess2013IVW,Lawlor2016,Hartwig2016downsides} takes the variant--exposure association from one genome-wide association study and the variant--outcome association from another, each as a published coefficient, and reports their ratio. With public repositories and automated software~\cite{Hemani2018MRBase}, anyone with internet access can run such an analysis~\cite{Burgess2024credibility}, and the literature has grown accordingly: from 61 PubMed-indexed papers in 2010, to over 18,000 PubMed records related to Mendelian randomization published since January 2024 to the time of writing~\cite{Tobias2024,Stender2024,Hemani2025growth}. Field reviews find recent studies to be two-sample summary-data analyses almost without exception~\cite{RosasChavez2025,Woolf2022}, and the design's quality problems have prompted calls for editors to reject stand-alone two-sample papers~\cite{Stender2024,Burgess2024credibility}.

This paper is about a different, substantive problem: what the ratio means. The ratio of a variant--outcome to a variant--exposure coefficient, the Wald or two-stage least squares estimand, has a causal interpretation when the outcome responds \emph{linearly} to the exposure. The textbook justification assumes a constant linear effect~\cite{Burgess2017review,Sanderson2022primer}. Identification results that let the effect vary between individuals, for a binary exposure and instrument by Wang and Tchetgen Tchetgen~\cite{WangTchetgen2018} and Cui and Tchetgen Tchetgen~\cite{CuiTchetgen2021}, and for continuous ones by Hartwig et al.~\cite{HartwigNOSH2023,HartwigTwoSample2026}, retain linearity within the individual. For a binary exposure it costs nothing: a function of a two-valued exposure is automatically linear in it, so the Wald ratio and its binary-exposure identification results are immune to the problems discussed here. For a continuous exposure it is a genuine restriction, doubtful for alcohol, adiposity, blood pressure, sleep, vitamin D and many biomarkers, which is why a literature on nonlinear Mendelian randomization exists~\cite{Burgess2014nonlinear,Silverwood2014,StaleyBurgess2017,Sulc2022,Tian2023}, one that requires individual-level data and has itself proved fragile~\cite{Burgess2023constant,Wade2023,Hamilton2024,Stender2024}.

We ask a prior question: when the dose-response is nonlinear and the standard two-sample pipeline is run anyway, what is estimated, and what do its diagnostics mean? We show that a single valid variant's ratio remains a causal quantity, a weighted average of the slope of the dose-response curve, through two separate lenses resting on distinct structural assumptions: no unmeasured common effect modifier (NUCEM), a restriction on how unmeasured heterogeneity in the outcome and in the first stage covary, and an exposure control function (ECoF), the existence of a function of exposure and instrument that controls confounding, of which the triangular model of Imbens and Newey~\cite{ImbensNewey2009} and the familiar location-shift first stage are examples. In both, the weight is set by how that variant moves the exposure distribution, and in both it is nonnegative exactly when the distribution functions of exposure at different genotypes do not cross, a comonotonicity condition that dispersive variants violate. Variants that move the exposure distribution differently therefore estimate different things, even in models where the dose-response curve itself is point identified from individual-level data: what two-sample Mendelian randomization loses is not identification but the meaning of the aggregate. We identify what individual-level exposure data make possible and argue that, for continuous exposures, such data should again be the standard. Proofs are in the Appendix; R code reproducing the illustration and all numerical checks is available from the author.

\section{The Wald estimand for a continuous exposure}
\label{sec:setting}

Let $A$ be a continuous exposure, $Y$ an outcome, $Z$ a variant (allele count) or score, and $U$ the unmeasured common causes of $A$ and $Y$; throughout, we ignore the presence of measured covariates although the extension is straightforward via conditioning. Write $Y(a)$ for the potential outcome under exposure $a$, $h(a,u)=\E\{Y(a)\mid U=u\}$ for the dose-response curve within a confounder stratum, $h'(a,u)=\partial_ah(a,u)$ for its slope, and $h'(a)=\E\{h'(a,U)\}$ for the population slope. Throughout, $\E_U$ and $\cov_U$ denote expectation and covariance over the distribution of $U$, $f(a\mid u,z)$ is the conditional density of $A$ given $U=u$ and $Z=z$, and $S(a\mid u,z)=\pr(A>a\mid U=u,Z=z)$ and $S(a\mid z)=\pr(A>a\mid Z=z)$ are conditional survival functions of the exposure. We assume the usual instrumental-variable conditions~\cite{HernanRobins2006}: $U$ suffices to control confounding of $A$ and $Y$, $Z$ affects $Y$ only through $A$, $Z$ is independent of $U$, and $Z$ is associated with $A$. These do not identify $h$ or any average of it~\cite{Swanson2018JASA}. Let $\gamma_Y=\cov(Z,Y)/\var(Z)$ and $\gamma_A=\cov(Z,A)/\var(Z)$ be the regression coefficients of outcome and exposure on the variant, the two quantities that two-sample Mendelian randomization estimates in separate samples. The Wald estimand is
\begin{equation}\label{eq:wald}
\beta_{\IV}=\frac{\cov(Z,Y)}{\cov(Z,A)}=\frac{\gamma_Y}{\gamma_A}.
\end{equation}
If $h(a,u)=\beta_0(u)+\beta a$ with one slope for everyone, $\beta_{\IV}=\beta$: the textbook case. We let $h(a,u)$ be an arbitrary smooth curve in each stratum and ask what \eqref{eq:wald} is then.

\section{What the ratio estimates when the dose-response is nonlinear}
\label{sec:nonlinear}

We reach a weighted-average-derivative interpretation of $\beta_{\IV}$ through two lenses that rest on different structural assumptions: the first keeps the latent confounder $U$ and restricts how heterogeneity in the outcome and in the first stage covary; the second replaces $U$ by an observable control function.

\subsection{Lens 1: no unmeasured common effect modifier}

Define the variant's \emph{weight} at dose $a$,
\[
g(a)=\cov\{Z,\ind(A>a)\},
\]
the covariance between the variant and the indicator that exposure exceeds $a$, and its within-stratum version $g(a,u)=\cov\{Z,\ind(A>a)\mid U=u\}$. Both effectively measure the strength of instrument-exposure association at dose $a$. Under the instrumental-variable conditions alone (Appendix~A),
\begin{equation}\label{eq:decomp}
\cov(Z,Y)=\int h'(a)\,g(a)\dd a+\int\cov_U\{h'(a,U),g(a,U)\}\dd a,\qquad \cov(Z,A)=\int g(a)\dd a.
\end{equation}
The first term averages the population slope with weight $g$; the second is residual confounding, the covariance across strata between the local slope and the local instrument strength, summed over dose levels. The first lens assumes it vanishes.

\begin{condition}[No unmeasured common effect modifier, NUCEM]\label{cond:gnosh}
$\int\cov_U\{h'(a,U),g(a,U)\}\dd a=0$.
\end{condition}

\noindent The name is that of the binary-exposure condition of Cui and Tchetgen Tchetgen~\cite{CuiTchetgen2021}, of which this is the continuous form: summed over doses, the confounder's shaping of the causal slope is uncorrelated with its shaping of the variant's effect on exceeding each dose.

\begin{result}\label{res:wad}
Under the instrumental-variable conditions and NUCEM,
\begin{equation}\label{eq:wad}
\beta_{\IV}=\frac{\int h'(a)\,g(a)\dd a}{\int g(a)\dd a}.
\end{equation}
The weight $g$ depends only on the joint distribution of $(Z,A)$.
\end{result}

\noindent The weight $g$ is the continuous-exposure form of the weight in Angrist and Imbens' average causal response theorem~\cite{AngristImbens1995,AngristGraddyImbens2000}. In general, it is only identified from individual-level genotype and exposure data.
The weight is also specific to the variant. A variant that acts only at low doses averages the slope over low doses; a variant that increases the \emph{spread} of exposure, as a variance quantitative trait locus does~\cite{Yang2012FTO,Wang2019vQTL}, has a weight that is negative below some dose and positive above it, so its ratio need not lie within the range of the slope at all (Section~\ref{sec:signs}).

\paragraph{Two examples.} \emph{No effect modification by the confounder}, $h(a,u)=h(a)+\nu(u)$, the continuous form of the outcome-side condition of Wang and Tchetgen Tchetgen~\cite{WangTchetgen2018}: NUCEM holds for every first stage, because $h'(a,u)=h'(a)$ makes each covariance in Condition~\ref{cond:gnosh} zero. \emph{Additive treatment density}, $f(a\mid u,z)=c(a,u)+\Lambda(a,z)$, so that the variant's effect on the exposure density is the same in every stratum~\cite{WangTchetgen2018,Michael2024,ChenZhangCui2026}: NUCEM holds for \emph{every} outcome model, because $g(a,u)$ does not depend on $u$. The condition holds when exposure is set through mutually exclusive variant-driven and confounder-driven routes (Fig.~\ref{fig:illus}), fails for the additive-error first stage, and under it the dose-response curve $h$ is itself point identified from individual-level data~\cite{Michael2024,ChenZhangCui2026}, while the Wald ratio reports only the $g$-weighted average of its slope.

\paragraph{The interpretation strengthens as the outcome model is restricted.} With a \emph{linear heterogeneous outcome},
\begin{equation}\label{eq:linear}
h(a,u)=\beta_0(u)+\beta_Y(u)\,a,
\end{equation}
each stratum has a single causal slope $\beta_Y(u)$ and the average causal effect is $\ACE=\E\{\beta_Y(U)\}$. Define the within-stratum instrument strength $\gamma_A(u)=\cov(A,Z\mid U=u)/\var(Z)$, with $\E\{\gamma_A(U)\}=\gamma_A$ because $Z\indep U$. Under \eqref{eq:linear} the integral in Condition~\ref{cond:gnosh} equals $\var(Z)$ times $\cov_U\{\beta_Y(U),\gamma_A(U)\}$, and \eqref{eq:decomp} reads (Appendix~A)
\begin{equation}\label{eq:noshgap}
\beta_{\IV}-\ACE=\frac{\cov_U\{\beta_Y(U),\gamma_A(U)\}}{\E\{\gamma_A(U)\}},
\end{equation}
so that NUCEM gives $\beta_{\IV}=\ACE$. This is the no simultaneous heterogeneity (NOSH) result of Hartwig et al.~\cite{HartwigNOSH2023}, who assume \eqref{eq:linear} (their Assumption 2) and that the unmeasured modifiers of the $Z$--$A$ effect are independent of the unmeasured modifiers of the $A$--$Y$ effect (Assumption 1), which under linearity makes $\beta_Y(U)\indep\gamma_A(U)$ and the covariance zero: NOSH is NUCEM plus linearity. For a binary exposure and instrument the result is that of Wang and Tchetgen Tchetgen~\cite{WangTchetgen2018} and Cui and Tchetgen Tchetgen~\cite{CuiTchetgen2021}. With a \emph{nonlinear additive outcome}, $h(a,u)=h(a)+\nu(u)$, linearity fails yet NUCEM holds for every first stage, and $\beta_{\IV}$ is the weighted average slope of the one causal curve $h$. Even without NUCEM, $\beta_{\IV}$ is a $g(a,u)$-weighted average of individual-level local slopes with unidentified weights (Appendix~A), a convex combination under the within-stratum form of the condition of Section~\ref{sec:signs}, the continuous analogue of Small et al.'s stochastic-monotonicity result~\cite{Small2017}, that the within-stratum exposure distribution does not cross across values of the instrument.

\subsection{Lens 2: an exposure control function}

The second lens replaces the latent confounder by one generated from the observed data. Suppose there is a function $T=T(A,Z)$ of exposure and instrument, an \emph{exposure control function} (ECoF), such that for every $a$
\begin{equation}\label{eq:cf}
Y(a)\indep(A,Z)\mid T,\qquad T\indep Z,\qquad A=\kappa(T,Z),
\end{equation}
with $\kappa$ differentiable in $z$, and, for identification of the curve, that the support of $T$ given $A=a$ equals the support of $T$ (common support). Then $m(a,t)=\E(Y\mid A=a,T=t)=\E\{Y(a)\mid T=t\}$ is a causal curve within each level of $T$, its average $\E_T\{m(a,T)\}=\E\{Y(a)\}$ is the dose-response curve, i.e. the g-formula where $T$ is sufficient to control for confounding, and both are point identified from individual-level data~\cite{ImbensNewey2009}; $T$ plays the role of $U$ in the instrumental-variable conditions, with the first stage degenerate within each level of $T$. Two models satisfy \eqref{eq:cf}. In the triangular model of Imbens and Newey~\cite{ImbensNewey2009}, $A=h(Z,\eta)$ with a scalar disturbance $\eta$ independent of $Z$ and $h$ strictly increasing in $\eta$, an ECoF is the conditional distribution function $T=F_{A\mid Z}(A\mid Z)$ and $\kappa(t,z)$ is the conditional quantile $Q_{A\mid Z}(t\mid z)$; this choice is an ECoF exactly when $\kappa(\cdot,z)$ orders the support of $T$ in the same way for every $z$; every ECoF is a $z$-dependent measure-preserving rearrangement of $F_{A\mid Z}(A\mid Z)$, one can be an ECoF when the Imbens--Newey choice is not, and the data cannot tell (Appendix~B gives a counter-example). In the location-shift first stage $A=\mu(Z)+V$ with $V\indep Z$, the model under which two-sample Mendelian randomization is usually motivated, $T=A-\mu(Z)$ and $\kappa(t,z)=\mu(z)+t$. The second is nested in the first, since $F_{A\mid Z}(A\mid Z)=F_V\{A-\mu(Z)\}$ is a monotone transformation of $A-\mu(Z)$.

\begin{result}\label{res:cf}
Under \eqref{eq:cf} and common support, with $w(z)=\cov\{Z,\ind(Z>z)\}\ge0$ and $f_Z$ the density of $Z$,
\begin{equation}\label{eq:cfwad}
\beta_{\IV}=\frac{\E\{\omega(Z,T)\,\partial_am(A,T)\}}{\E\{\omega(Z,T)\}},\qquad
\omega(z,t)=\frac{\partial\kappa(t,z)}{\partial z}\,\frac{w(z)}{f_Z(z)} .
\end{equation}
The weight is identified from exposure and genotype data once $T$ is.
\end{result}

\noindent The proof (Appendix~B) leverages integration by parts within each level of $T$: the Wald ratio is a weighted average of the control-function slope at each person's own exposure, weighted by the person's responsiveness $\partial_z\kappa$ to the instrument and a factor depending only on the instrument's distribution. Under the location shift the weight is $\omega(z)=\mu'(z)w(z)/f_Z(z)$, a function of the instrument alone. NUCEM plays no part here and generally fails, because within each level of $T$ the exposure is a deterministic function of $Z$ (Appendix~B); again the curve is identified from individual-level data, and the Wald ratio reports one responsiveness-weighted functional of it.

Again the interpretation strengthens with the outcome model. Under the \emph{proportional perturbation} condition that $\omega$ is constant, which holds for a normally distributed instrument such as a polygenic score with a location-shift first stage with linear $\mu$, \eqref{eq:cfwad} equals $\E\{\partial_am(A,T)\}$, the average derivative of Imbens and Newey: the mean of each individual's causal slope at the individual's own exposure, obtained with no linearity. Appendix~B restates the condition directly on the distribution of exposure given instrument and confounder. With a \emph{linear heterogeneous outcome}, $\E\{Y(a)\mid T=t\}=\beta_0(t)+\beta_Y(t)a$,
\begin{equation}\label{eq:cflinear}
\beta_{\IV}=\frac{\E\{\beta_Y(T)\,\gamma_A(T)\}}{\E\{\gamma_A(T)\}},\qquad \gamma_A(t)=\frac{\cov(A,Z\mid T=t)}{\var(Z)},
\end{equation}
which is \eqref{eq:noshgap} with $T$ for $U$: $\beta_{\IV}=\ACE$ exactly when NUCEM holds with the control function as the confounder, automatically under the location shift, where $\gamma_A(t)$ is constant, but not in the Imbens--Newey model when responsiveness $\partial_zh(Z,\eta)$ covaries with $\beta_Y$. With a \emph{nonlinear additive outcome}, $\E\{Y(a)\mid T\}=h(a)+\nu(T)$, \eqref{eq:cfwad} collapses to \eqref{eq:wad} with the same identified weight $g(a)=\cov\{Z,\ind(A>a)\}$ (Appendix~B): the two lenses agree. For an allele count $Z\in\{0,1,2\}$ under the location shift with $\mu(z)=\mu_0+\beta z$ and effect-allele frequency $p$, writing $H(m)=\E\{h(m+V,U)\}$ and $H''$ for its curvature at the population mean,
\begin{equation}\label{eq:benign}
\beta_{\IV}=\E\{h'(A,U)\}+H''\,\beta\,(\tfrac12-p)+O(\beta^2).
\end{equation}
Every variant therefore targets the same average slope up to a term proportional to its effect size, allele frequency and the curvature, small for typical per-allele effects. We call this the \emph{benign case}, benign because the first stage acts identically on everyone.

\subsection{Nonnegative weights: comonotonicity and non-crossing distribution functions}
\label{sec:signs}

A weighted average lies within the range of what it averages only if its weights are nonnegative; under both lenses this is the same condition on the first stage alone.

\begin{result}\label{res:signs}
(i) Under lens 1, $g(a)=\cov\{Z,S(a\mid Z)\}\ge0$ for every $a$ whenever $z\mapsto S(a\mid z)$ is nondecreasing for every $a$, that is, whenever $Z$ and $S(a\mid Z)$ are comonotone; if $S(a_0\mid\cdot)$ is not monotone, some instrument distribution makes $g(a_0)<0$. (ii) Under lens 2, $\omega\ge0$ if and only if $\partial_z\kappa\ge0$; in the Imbens--Newey model this holds for every $t$ if and only if $S(a\mid z)$ is nondecreasing in $z$ for every $a$, and under the location shift if and only if $\mu$ is nondecreasing.
\end{result}

\noindent (Appendix~G.) Both lenses therefore ask that higher values of the variant shift the exposure distribution upward in the sense of first-order stochastic dominance: the exposure distribution functions at the three genotypes, or along a score, \emph{do not cross}. When the condition holds, $\beta_{\IV}$ lies within the range of $h'$ under lens 1 and of $\partial_am$ under lens 2. It is the continuous analogue of no defiers~\cite{AngristImbens1995,Small2017} and of monotonicity in the instrument in the triangular model, which Imbens and Newey do not assume. Sufficient primitives are a location shift with nondecreasing $\mu$, a monotone likelihood ratio in $(a,z)$, or a transformation model $\tau(A)=\mu(Z)+\xi$ with $\tau$ increasing and $\xi\indep Z$ (Appendix~G).

When the distribution functions cross, the weights take both signs under both lenses. The leading case is a location--scale first stage $A=\mu(Z)+\sigma(Z)\eta$ with $\sigma'>0$, the structure attributed to variance loci: $\partial_z\kappa=\mu'+\sigma'\eta$ is negative for low $\eta$, and with an unbounded error any scale effect, however small, produces crossing, with negative weight below the crossing point and positive above it (Appendix~G); variant 4 of Fig.~\ref{fig:illus} is such a case, and its ratio exceeds every individual effect. Crossing is invisible in the variant--exposure coefficient $\gamma_A$, which integrates $g$ and can be positive and large while $g$ changes sign. It is checkable from individual-level exposure and genotype data by comparing empirical distribution functions across genotypes or by quantile regression of exposure on allele count (Appendix~G); a summary-data analysis cannot check it.

\subsection{The two lenses compared}

Table~\ref{tab:routes} provides the summary. Neither lens implies the other: NUCEM restricts latent heterogeneity without restricting the first stage, while an ECoF makes the first stage degenerate within $T$ and generally violates NUCEM.

\begin{table}[!htbp]\centering\footnotesize\setlength{\tabcolsep}{4pt}
\caption{Two lenses under which the Wald ratio is a weighted average derivative ($g(a)=\cov\{Z,\ind(A>a)\}$; $\omega$ as in \eqref{eq:cfwad}; $m(a,t)=\E\{Y(a)\mid T=t\}$). Under both lenses the weight is nonnegative exactly when the distribution functions of exposure given the variant do not cross.}
\label{tab:routes}
\begin{tabular}{p{3.5cm}p{3.6cm}p{3.6cm}p{3.5cm}}
\toprule
 & Unrestricted outcome & Linear heterogeneous outcome \eqref{eq:linear} & Nonlinear additive outcome $h(a)+\nu(u)$\\
\midrule
\textbf{Lens 1: NUCEM} with latent $U$. Examples: no effect modification; additive treatment density. &
$\int h'g/\int g$; weight identified from $(A,Z)$; proper iff $S(a\mid z)$ nondecreasing in $z$ &
ACE; NUCEM $+$ linearity is NOSH &
$\int h'g/\int g$ with one curve $h$; NUCEM holds, NOSH fails\\[4pt]
\textbf{Lens 2: ECoF} $T(A,Z)$. Examples: Imbens--Newey, $T=F_{A\mid Z}(A\mid Z)$; location shift, $T=A-\mu(Z)$ ($\subset$ Imbens--Newey). &
$\E\{\omega\,\partial_am\}/\E\{\omega\}$; proper iff $\partial_z\kappa\ge0$, the same non-crossing; $=\E\{\partial_am(A,T)\}$ under proportional perturbation &
$\E\{\beta_Y(T)\gamma_A(T)\}/\E\{\gamma_A(T)\}$; ACE iff NUCEM with $T$ as confounder (automatic under location shift) &
$\int h'g/\int g$, identical to lens 1; NOSH fails\\
\bottomrule
\end{tabular}
\end{table}

\paragraph{Implication for practice.} A single valid variant's ratio is causal under interpretable conditions even when the dose-response is nonlinear, and either lens says which quantity it is. But that quantity belongs to the variant as much as to the exposure--outcome relation, and it can be read off only if one can see how the variant moves the exposure distribution, which summary statistics do not show.

\section{Consequences for the summary-data pipeline}
\label{sec:pipeline}

Take $L$ mutually independent valid variants $G_1,\dots,G_L$, each with allele frequency $p_j$, so $\var(G_j)=2p_j(1-p_j)$, each satisfying NUCEM (as in either lens-1 example). Write $\gamma_{Y,j}$ and $\gamma_{A,j}$ for the outcome and exposure coefficients on $G_j$, $\hat\gamma_{Y,j}$, $\hat\gamma_{A,j}$ for their estimates from the two samples, and $\hat\theta_j=\hat\gamma_{Y,j}/\hat\gamma_{A,j}$ for the estimated ratio, whose population value is, by Result~\ref{res:wad},
\begin{equation}\label{eq:thetaj}
\theta_j=\frac{\gamma_{Y,j}}{\gamma_{A,j}}=\frac{\int h'(a)\,g_j(a)\dd a}{\int g_j(a)\dd a},\qquad g_j(a)=\cov\{G_j,\ind(A>a)\}.
\end{equation}
Under a linear dose-response all $\theta_j$ equal the ACE~\cite{HartwigTwoSample2026}; under a nonlinear one they are $L$ different functionals of one curve, even where that curve is point identified from individual-level data. The same conclusions hold under lens 2 with responsiveness-weighted averages in place of \eqref{eq:thetaj}. Details for each estimator are in Appendix~C; the conclusions are these.

The \emph{inverse-variance weighted} estimand~\cite{Burgess2013IVW}, $\theta_{\mathrm{IVW}}=\sum_j\lambda_j\theta_j$ with precision weights $\lambda_j$ proportional to $\gamma_{A,j}^2\var(G_j)$, is a precision-weighted mixture of the $\theta_j$, hence a weighted average slope whose weight is a mixture of the variant-specific weights. It is causal, and proper if every $g_j\ge0$, but it is not a parameter of the exposure--outcome relation: it changes with the variants selected, the $p$-value threshold and clumping that selected them, and the sample sizes that set the precision weights. Two analysts with different valid instrument sets estimate different parameters and, at current sample sizes, will confidently disagree.

\emph{Cochran's $Q$}~\cite{Bowden2017framework}, the precision-weighted dispersion of the $\hat\theta_j$ about $\hat\theta_{\mathrm{IVW}}$, has noncentrality $\sum_j(\theta_j-\theta_{\mathrm{IVW}})^2/\mathrm{se}(\hat\theta_j)^2$, which grows linearly with the outcome sample size $n_Y$. In the benign case the deviations are small, but with hundreds of variants and samples in the millions, $I^2$ values of tens of percent arise from curvature alone. In the non-benign case the deviations are of the order of the spread of $h'$ and $Q$ is large with a handful of variants. The heterogeneity is real, but it is heterogeneity of estimands, not evidence against the instruments.

\begin{figure}[!htbp]\centering
\includegraphics[width=\textwidth]{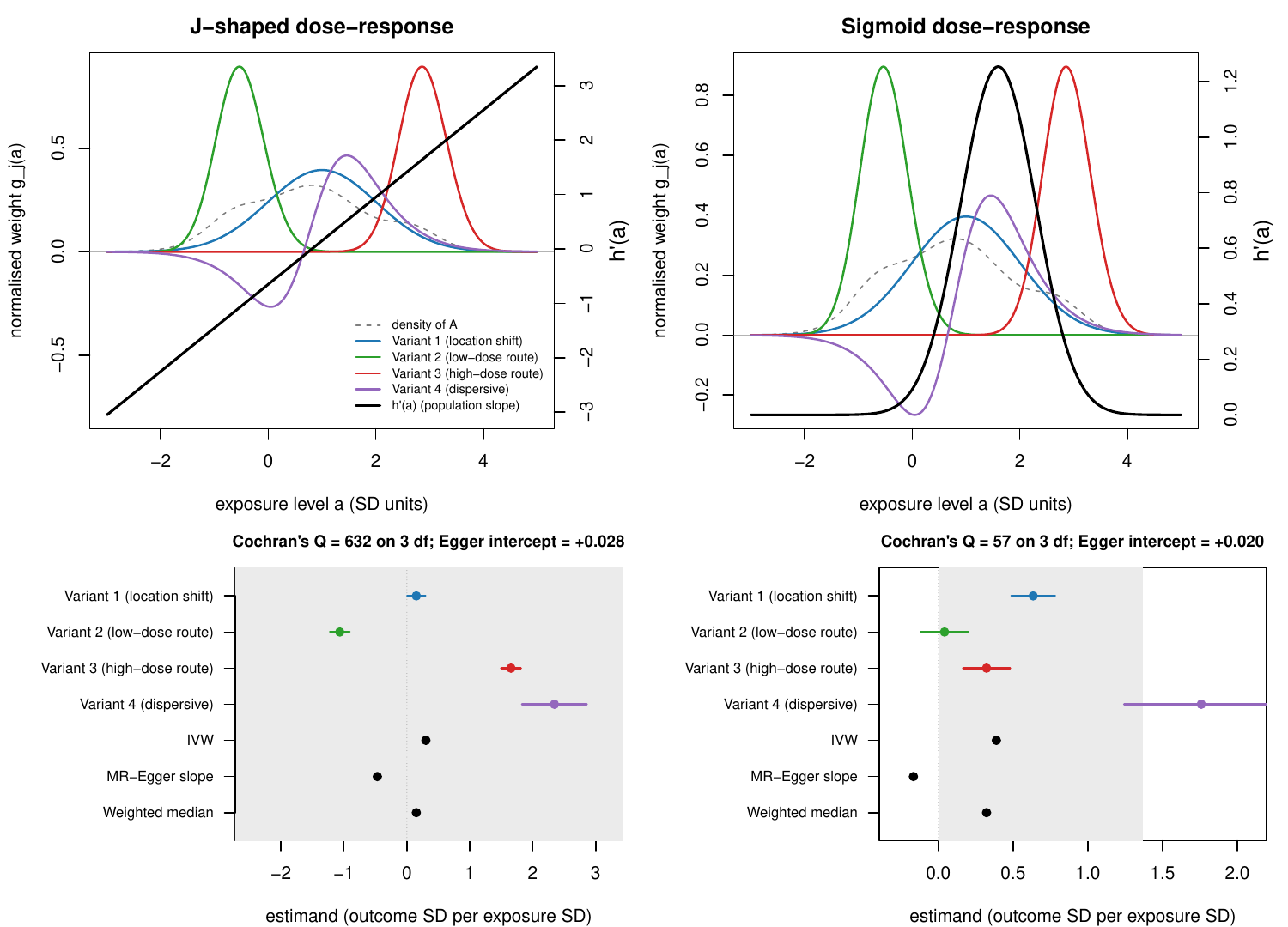}
\caption{Four valid variants, one nonlinear dose-response, and the summary-data pipeline. Exposure is set through mutually exclusive routes, so the additive-density condition and NUCEM hold and each Wald ratio is the weighted average slope \eqref{eq:thetaj}. Variant 1 shifts the whole exposure distribution; variant 2 acts in a low-dose route; variant 3 in a high-dose route; variant 4 increases dispersion. \emph{Top:} population slope $h'(a)$ (black) and each variant's normalized weight for a J-shaped and a sigmoid dose-response; variant 4's weight is negative below the center. \emph{Bottom:} variant-specific estimands with 95\% intervals reflecting an outcome sample of 300,000, the inverse-variance weighted, MR-Egger and weighted-median estimands, and (shaded) the range of individual-level slopes. For the sigmoid curve variant 4's estimand, 1.76, exceeds every individual effect ($\le1.37$); $Q$ rejects homogeneity decisively in both cases; MR-Egger reports a nonzero intercept and, for the J-shaped curve, a slope of the wrong sign. Parameters in Appendix~F.}
\label{fig:illus}
\end{figure}

\emph{MR-Egger}~\cite{Bowden2015Egger} assumes that pleiotropic effects are independent of instrument strength (InSIDE). Under nonlinearity $\gamma_{Y,j}=\theta_j\gamma_{A,j}$, so relative to any common value $\bar\theta$ the ``pleiotropic'' part $(\theta_j-\bar\theta)\gamma_{A,j}$ is \emph{proportional} to instrument strength by construction. In the benign case, \eqref{eq:benign} makes the variant--outcome coefficient approximately quadratic in the variant--exposure coefficient, which a line fits with a nonzero intercept: curvature is reported as directional pleiotropy. In the non-benign case the slope can have the wrong sign (Fig.~\ref{fig:illus}).

The \emph{weighted median}~\cite{Bowden2016median} converges to the estimand of one variant, the median one; its ``50\% valid'' guarantee is vacuous because every variant is valid. The \emph{mode}~\cite{Hartwig2017mode} returns the estimand shared by the largest cluster, the variants that move the exposure in the most common way. \emph{Outlier removal}~\cite{Verbanck2018} deletes the variants whose weights sit where the slope is unusual, which is selection on the estimand. If some $g_j$ change sign, that is, if some variant's exposure distribution functions cross, neither the $\theta_j$ nor any aggregate need lie within the range of individual effects.

\paragraph{Illustration.}
Figure~\ref{fig:illus} and Table~\ref{tab:illus} show four valid variants acting through mutually exclusive exposure routes, without pleiotropy, weak instruments or sample overlap. For a J-shaped dose-response the low-dose and high-dose variants give $-1.07$ and $+1.65$ outcome SD per exposure SD, MR-Egger a slope of $-0.47$ with a nonzero intercept, and $Q=632$ on 3 df; for a sigmoid dose-response whose slope is at most $1.37$ for every individual, the dispersive variant's ratio is $1.76$. These are population values.

\begin{table}[!htbp]\centering\footnotesize\setlength{\tabcolsep}{4pt}
\caption{Population values in the illustration of Fig.~\ref{fig:illus} (outcome SD per exposure SD; $Q$ for an outcome sample of 300,000).}
\label{tab:illus}
\begin{tabular}{lcccccc}
\toprule
 & $\gamma_{A,j}$ & \multicolumn{2}{c}{J-shaped} & & \multicolumn{2}{c}{Sigmoid, $h'\in[0,1.37]$}\\
\cmidrule{3-4}\cmidrule{6-7}
 & & $\theta_j$ & negative weight & & $\theta_j$ & negative weight\\
\midrule
Variant 1 (location shift) & 0.038 & $+0.15$ & 0\% & & $+0.63$ & 0\%\\
Variant 2 (low-dose route) & 0.040 & $-1.07$ & 0\% & & $+0.04$ & 0\%\\
Variant 3 (high-dose route) & 0.040 & $+1.65$ & 0\% & & $+0.32$ & 0\%\\
Variant 4 (dispersive) & 0.010 & $+2.34$ & 33\% & & $+1.76$ & 33\%\\
\midrule
Inverse-variance weighted & & $+0.30$ & & & $+0.39$ & \\
MR-Egger slope (intercept) & & $-0.47$ ($+0.028$) & & & $-0.17$ ($+0.020$) & \\
Weighted median & & $+0.15$ & & & $+0.32$ & \\
Cochran's $Q$ (3 df) & & 632 & & & 57 & \\
\bottomrule
\end{tabular}
\end{table}

\paragraph{Is the non-benign case realistic?}
It requires variants that move the exposure distribution differently, which are well documented: variance loci for adiposity and other traits, generally read as gene--environment interaction~\cite{Yang2012FTO,Wang2019vQTL}; the \textit{ALDH2} and \textit{ADH1B} variants for alcohol, which act only among drinkers and whose effect on intake differs sharply by sex and setting~\cite{Chen2008ALDH2,Millwood2019}; smoking-heaviness variants that act only in smokers; and variants with effects modified by sex, age, diet or ancestry~\cite{Spiller2019}. Each produces its own weight. The benign expansion \eqref{eq:benign} describes a best case.

\section{Summary statistics cannot attribute the heterogeneity}
\label{sec:diagnostics}

Every pipeline diagnostic tests whether the $\hat\theta_j$ are homogeneous. Nonlinearity makes them heterogeneous with valid instruments. Can summary statistics tell which?

\begin{proposition}\label{prop:equiv}
Any collection of summary statistics $\{(\gamma_{A,j},\gamma_{Y,j})\}$ with $\gamma_{A,j}\neq0$, with any allele frequencies, is generated exactly both by (i) valid instruments satisfying the additive-density condition, with nonnegative weights and a nonlinear dose-response, and by (ii) a linear constant-effect dose-response with a direct effect of each variant on the outcome.
\end{proposition}

\noindent The proof (Appendix~D) is constructive. The two processes differ in the conditional distribution of exposure given genotype and agree on everything a summary-data analysis sees. No function of the summary statistics, including $Q$, the MR-Egger intercept, funnel asymmetry, MR-PRESSO and leave-one-out analyses, separates them. The InSIDE, plurality and zero-modal-pleiotropy assumptions are, when nonlinearity is possible, assumptions about the first stage dressed as assumptions about pleiotropy, and untestable against it.

One qualification is needed. If the exposure study reported each variant's association with a few transforms of the exposure ($A^2$, $A^3$, splines, or threshold indicators $\ind(A>a)$), the weights $g_j$ would become available and the next section would apply. These statistics are cheap, estimable at scale~\cite{Wang2019vQTL}, and almost never reported.

\section{What individual-level exposure data buys: hybrid designs}
\label{sec:hybrid}

``Two-sample'' sometimes means individual-level genotype and exposure data in a biobank and summary-level outcome data from a consortium (design $H_1$), or the mirror image, individual-level outcome and summary-level exposure ($H_2$). The two are not symmetric.

Under $H_1$, each weight $g_j(a)=\cov\{G_j,\ind(A>a)\}$ is estimable, and so are the ECoF and responsiveness of lens 2, so crossing can be checked for every variant and the doses its ratio averages over reported. Moreover, for any basis $\phi_1,\dots,\phi_K$ with antiderivatives $\Phi_k$, $\int\phi_kg_j=\cov\{G_j,\Phi_k(A)\}$, so if $h'(a)=\sum_kc_k\phi_k(a)$ and NUCEM holds for each variant,
\begin{equation}\label{eq:hybrid}
\gamma_{Y,j}\var(G_j)=\sum_{k=1}^Kc_k\,\cov\{G_j,\Phi_k(A)\},\qquad j=1,\dots,L,
\end{equation}
a linear system whose right side comes from the exposure sample and whose left side is the outcome summary statistic scaled by the allele variance (Appendix~E). The slope of the dose-response curve is identified on the span of the basis if and only if the variants' associations with $\Phi_1(A),\dots,\Phi_K(A)$ are not proportional across variants, that is, if and only if the weights differ in shape: the heterogeneity that corrupts inverse-variance weighting is exactly what identifies the curve. In the benign case the matrix has rank one and only the average slope is identified, which is what inverse-variance weighting then estimates. The design delivers either an interpretable pooled estimate or an estimate of the curve, and says which; with more variants than basis functions, the residual of \eqref{eq:hybrid} assesses the joint hypothesis of valid instruments and a slope in the span. In the illustration, variants 1--3 recover a linear slope exactly. Relation \eqref{eq:hybrid} is a two-sample form of polynomial Mendelian randomization~\cite{Sulc2022} that needs no stratification on the exposure and so avoids the biases that have troubled stratified nonlinear methods~\cite{Burgess2023constant,Hamilton2024}.

Under $H_2$ the weights are properties of the exposure sample and cannot be recovered from $\gamma_{A,j}$; unless the exposure study publishes the distributional statistics just described, $H_2$ is no better than a pure summary-data design. The sample that must be individual-level is the exposure sample. With individual-level data on all three variables, in one sample or two, everything above is available, and more: non-crossing, which decides whether a variant's weight is nonnegative in either lens, can be checked by quantile regression of exposure on genotype (Appendix~G), and its within-stratum form probed within observed covariate strata; the control-function curve $m(a,t)$ and its average can be estimated directly~\cite{ImbensNewey2009}; and the curve under the additive-density condition can be estimated by semiparametric or nonparametric instrumental-variable methods~\cite{ChenZhangCui2026,NeweyPowell2003}.

\section{Discussion}

For a continuous exposure with a nonlinear effect, a single valid variant's Wald ratio is a causal quantity through either of two lenses, no unmeasured common effect modifier, a restriction on the covariation of latent heterogeneities in the outcome and the first stage, or an exposure control function, under which the confounder is observable and the weights are each person's responsiveness to the instrument. In both the quantity is a weighted average of the slope of the dose-response, proper exactly when the exposure distribution functions at different genotypes do not cross. But the quantity belongs to the variant as much as to the exposure--outcome relation, and that is what the multi-variant machinery of two-sample Mendelian randomization cannot appropriately account for, or absorb. Each variant estimates its own functional of the dose-response curve; the pooled estimand changes with the instrument set; heterogeneity statistics and the MR-Egger intercept respond to nonlinearity exactly as to pleiotropy; the median and mode return one variant's estimand without saying which; and none of this is detectable from summary statistics. The loss is not one of identification: in the additive-density and ECoF models the dose-response curve is point identified from individual-level data, and under a hybrid design its slope is identified from the pipeline's own outcome statistics together with individual-level exposure data. What the summary-data aggregate lacks is a meaning, not a target. A warning follows for NOSH~\cite{HartwigNOSH2023}, the most careful existing justification of the ratio for a continuous exposure: it presumes linearity, which is unlikely for many exposures, and it is the linearity, not the independence of heterogeneities, that makes every variant estimate the same thing. NUCEM and an ECoF address the resulting difficulty of interpreting two-stage least squares to some extent, by saying what a single ratio is without linearity, but they do not rescue the aggregate.

There is a benign case, which we have stated: when every variant shifts every person's exposure equally, the pipeline estimates, approximately, the average of individual slopes at individual doses. But that is a statement about the first stage, and the first stage is precisely where genetic epidemiology has documented pervasive heterogeneity; the documentation of variance loci is evidence of crossing distribution functions. For such instruments the non-benign case prevails.

Stender et al.~\cite{Stender2024} advise editors to reject stand-alone two-sample papers because they add nothing to what can be looked up online. Our results add a substantive reason for continuous exposures: what is looked up need not have a causal meaning. We recommend, first, that for a continuous exposure a pooled two-sample estimate be reported with an explicit, defended linearity assumption, as STROBE-MR already requires for the core assumptions~\cite{Skrivankova2021}. Second, that heterogeneity statistics and the MR-Egger intercept not be read as pleiotropy without an argument that the dose-response is linear or the first stages additive; ``robust'' sensitivity analyses offer no protection against nonlinearity. Third, that exposure consortia release variant associations with a few transforms or thresholds of the exposure, at negligible cost, so that analysts can recover the weights and check for crossing. Fourth, and chiefly, that individual-level data be reprioritized: a hybrid design with individual-level exposure data recovers the weights and, under NUCEM, the dose-response slope, and a one-sample design permits direct examination of the genotype--exposure relation and control-function estimation of the curve. The advantages that made two-sample designs attractive, larger samples, less winner's curse, weak-instrument bias toward the null~\cite{Pierce2013,Burgess2016overlap,Minelli2021}, were established under a linear model in which every variant estimates the same thing. Once that is in doubt, the economy of summary statistics is bought at the price of not knowing what is being estimated.

A final observation. The recent controversy over nonlinear Mendelian randomization~\cite{Burgess2023constant,Wade2023,Hamilton2024,Stender2024} concerns bias in stratification methods when the genetic effect on exposure is not constant across strata. That is the same first-stage heterogeneity that, here, makes variant-specific weights differ and summary-data aggregates uninterpretable. The two problems have one source, and the remedy for both begins with examining the genotype--exposure relation in individual-level data rather than reducing it to a single coefficient. The two stratification methods of nonlinear Mendelian randomization are, in fact, stratifications on the two exposure control functions used here. The residual method~\cite{Burgess2014nonlinear} stratifies on $A-\mu(Z)$, an ECoF only under the location shift, which is the constant-genetic-effect assumption; the rank-preserving assumption of the doubly ranked method~\cite{Tian2023,Burgess2023constant}, that the ordering of counterfactual exposures does not change with the instrument, is exactly the $z$-invariant ordering under which $F_{A\mid Z}(A\mid Z)$ is an ECoF, and its strata are coarsened levels of that function. Result~\ref{res:cf} applied within strata shows that each stratum-specific ratio is a responsiveness-weighted average of the causal slope over the exposures its members take across instrument values, proper only when the conditional quantile curves do not cross; the heterogeneity in genetic associations across doubly ranked strata documented in ref.~\cite{Burgess2023constant} is heterogeneity in that responsiveness.

\appendix
\setcounter{section}{0}
\renewcommand{\thesection}{\Alph{section}}
\section*{Appendix}
\addcontentsline{toc}{section}{Appendix}
\renewcommand{\theequation}{\Alph{section}.\arabic{equation}}

\noindent Notation is that of the main text. Two further objects are used below. With $f(a\mid u)$ the conditional density of $A$ given $U=u$ and $\rho(a,u)=g(a,u)/f(a\mid u)$, the instrumental-variable conditions alone give the representation
\begin{equation}\label{S:owndose}\tag{A.0}
\beta_{\IV}=\frac{\E\{\rho(A,U)\,h'(A,U)\}}{\E\{\rho(A,U)\}},
\end{equation}
proved in Appendix~A; $\rho\equiv$ constant, equivalently $g(a,u)=\cov(Z,A)f(a\mid u)$, is the \emph{proportional perturbation} condition of Section~\ref{sec:nonlinear}. The condition for nonnegative weights under both lenses (Result~\ref{res:signs}) is that $S(a\mid z)$ be nondecreasing in $z$ for every $a$; its within-stratum form is $S(a\mid u,z)$ nondecreasing in $z$. The illustration referred to below is that of Figure~\ref{fig:illus} and Table~\ref{tab:illus}, with parameters in Appendix~F.

\renewcommand{\theequation}{\Alph{section}.\arabic{equation}}
\section{Lens 1: proofs of the identities \eqref{eq:noshgap}, \eqref{eq:decomp}, \eqref{eq:wad} and \eqref{S:owndose}, and the examples}
\label{app:proofs}
\setcounter{equation}{0}

\paragraph{Identity \eqref{eq:noshgap}.}
Under \eqref{eq:linear} and the instrumental-variable conditions, $\E(Y\mid A,U,Z)=h(A,U)=\beta_0(U)+\beta_Y(U)A$. With $\bar z=\E(Z)$ and $\E(Z\mid U)=\bar z$,
\[
\cov(Z,Y)=\E[(Z-\bar z)\{\beta_0(U)+\beta_Y(U)A\}]=\E[\beta_Y(U)\E\{(Z-\bar z)A\mid U\}]=\E\{\beta_Y(U)\cov(A,Z\mid U)\}.
\]
Divide by $\cov(A,Z)=\E\{\cov(A,Z\mid U)\}=\var(Z)\E\{\gamma_A(U)\}$ and subtract the average effect. Under Assumption 1, $\beta_Y(U)\indep\gamma_A(U)$, so the covariance is zero.

\paragraph{Decomposition \eqref{eq:decomp}.}
By exclusion, latent ignorability and consistency, $\E(Y\mid A=a,U=u,Z=z)=h(a,u)$, so $\cov(Z,Y)=\E[(Z-\bar z)h(A,U)]$. Conditioning on $U=u$ and using $Z\indep U$, the inner expectation is $\int h(a,u)M(a,u)\dd a$ with $M(a,u)=\cov\{Z,f(a\mid u,Z)\}$. Since $\int_a^\infty M(s,u)\dd s=\cov\{Z,S(a\mid u,Z)\}=g(a,u)$, we have $M=-\partial_ag$ and $g(\pm\infty,u)=0$; integrating by parts, $\int h(a,u)M(a,u)\dd a=\int h'(a,u)g(a,u)\dd a$. Take expectations over $U$ and split $\E_U\{h'g\}=h'\E_U\{g\}+\cov_U\{h',g\}$. Setting $h(a,u)\equiv a$ gives the denominator.

\paragraph{Representations \eqref{eq:wad} and \eqref{S:owndose}; nonnegativity of the weight.}
Under NUCEM the second term of \eqref{eq:decomp} vanishes; divide. If $S(a\mid z)=\pr(A>a\mid Z=z)$ is nondecreasing in $z$, $g(a)=\cov\{Z,S(a\mid Z)\}\ge0$ as the covariance of comonotone functions of $Z$; conversely, if $S(a_0\mid z)$ is not monotone, an instrument distribution concentrated on a reversed pair of values gives $g(a_0)<0$, so monotonicity is necessary for nonnegativity to hold for every instrument distribution. With $g\ge0$, \eqref{eq:wad} is a convex combination of values of $h'$. Before the covariance split, $\cov(Z,Y)=\iint h'(a,u)g(a,u)\dd P(u)\dd a=\E\{\rho(A,U)h'(A,U)\}$ and $\cov(Z,A)=\E\{\rho(A,U)\}$, which is \eqref{S:owndose}: without NUCEM, $\beta_{\IV}$ is a $g(a,u)$-weighted average of individual-level slopes, and under the within-stratum non-crossing condition of Appendix~G, $g(a,u)\ge0$, it is a convex combination of values of $h'(a,u)$ with unidentified weights.

\paragraph{No effect modification.} $h'(a,u)=h'(a)$, so $\cov_U\{h'(a,U),g(a,U)\}=0$ for each $a$ and every first stage.

\paragraph{Additive treatment density.} Integrating $f(a\mid u,z)=c(a,u)+\Lambda(a,z)$ over $(a,\infty)$ gives $S(a\mid u,z)=S_c(a,u)+S_\Lambda(a,z)$, so $g(a,u)=\cov\{Z,S_\Lambda(a,Z)\}$ is free of $u$ and NUCEM holds for any $h$. Integrating over $a$, $\cov(A,Z\mid U=u)$ is also free of $u$, so under \eqref{eq:linear} Assumption 1's covariance condition holds automatically and $\beta_{\IV}=\ACE$; with $h$ linear, $h'(a)=\E\{\beta_Y(U)\}$ cancels in \eqref{eq:wad}. The additive-error first stage $A=\alpha+\beta Z+\gamma U+\varepsilon$ with $\varepsilon\indep(Z,U)$ and density $f_\varepsilon$ has conditional density $f(a\mid u,z)=f_\varepsilon(a-\alpha-\beta z-\gamma u)$, whose $z$-derivative depends on $u$, so it violates the condition. A mixture over mutually exclusive routes, $f(a\mid u,z)=\pi_z(z)f_z(a\mid z)+\pi_u(u)f_u(a\mid u)+\{1-\pi_z(z)-\pi_u(u)\}f_0(a)$, satisfies it because the $z$-derivative is free of $u$.

\paragraph{The strengthening ladder and NOSH.} With an unrestricted outcome and NUCEM, \eqref{eq:wad} holds with the variant's own weight. Under \eqref{eq:linear}, $h'(a,u)=\beta_Y(u)$ and $\int g(a,u)\dd a=\var(Z)\gamma_A(u)$, so the NUCEM integral is $\var(Z)\cov_U\{\beta_Y(U),\gamma_A(U)\}$ and NUCEM is the covariance condition of \eqref{eq:noshgap}; then $\beta_{\IV}=\ACE$. Hartwig et al.\ assume \eqref{eq:linear} (their Assumption 2) and that the unmeasured modifiers of the $Z$--$A$ effect are independent of the unmeasured modifiers of the $A$--$Y$ effect (Assumption 1); under Assumption 2 these imply $\beta_Y(U)\indep\gamma_A(U)$, since the two depend on $U$ only through modifiers of the $A$--$Y$ and of the $Z$--$A$ effect respectively, so the covariance is zero and $\beta_{\IV}=\ACE$, their Theorem 1; NOSH is NUCEM plus linearity. Under $h(a,u)=h(a)+\nu(u)$ with $h$ nonlinear, linearity fails, NUCEM holds, and \eqref{eq:wad} is a weighted average of the slope of the single curve $h$.

\section{Lens 2: the exposure-control-function representation \eqref{eq:cfwad} and its consequences}
\label{app:cf}
\setcounter{equation}{0}

\paragraph{Proof of \eqref{eq:cfwad}.}
Under \eqref{eq:cf}, $Y(a)\indep A\mid T$ gives $\E(Y\mid A=a,T=t)=\E\{Y(a)\mid T=t\}=m(a,t)$, and $T\indep Z$ with $Y(a)\indep Z\mid T$ makes $T$ a confounder satisfying conditions (ii)--(iii). Common support gives $\E\{Y(a)\}=\E_T\{m(a,T)\}$ with $m(a,\cdot)$ identified on the whole support of $T$, which is Theorem 3 of Imbens and Newey~\cite{ImbensNewey2009} for the identity transformation of $Y$. Within $T=t$ the exposure is $A=\kappa(t,Z)$, so by $T\indep Z$, $\cov(Z,Y)=\E_T[\cov\{Z,m(\kappa(T,Z),T)\mid T\}]$ and $\cov(Z,A)=\E_T[\cov\{Z,\kappa(T,Z)\mid T\}]$. By Yitzhaki's identity~\cite{Yitzhaki1996}, for continuous $Z$ with density $f_Z$ and any differentiable $\varphi$, $\cov\{Z,\varphi(Z)\}=\int\varphi'(z)w(z)\dd z$ with $w(z)=\cov\{Z,\ind(Z>z)\}\ge0$; applied for fixed $t$ to $\varphi(z)=m(\kappa(t,z),t)$, with $\varphi'=\partial_am\cdot\partial_z\kappa$, and to $\varphi=\kappa(t,\cdot)$,
\begin{equation}\label{eq:cfproof}
\beta_{\IV}=\frac{\E_T\int\partial_am(\kappa(T,z),T)\,\partial_z\kappa(T,z)\,w(z)\dd z}{\E_T\int\partial_z\kappa(T,z)\,w(z)\dd z}
=\frac{\E\{\omega(Z,T)\,\partial_am(A,T)\}}{\E\{\omega(Z,T)\}},
\end{equation}
which is \eqref{eq:cfwad}. For discrete $Z$ the identity reads $\cov\{Z,\varphi(Z)\}=\sum_k\{\varphi(z_{k+1})-\varphi(z_k)\}w_k$ with $w_k=\cov\{Z,\ind(Z>z_k)\}$, and the derivative in $z$ becomes a difference. Since $w\ge0$, $\omega\ge0$ if and only if $\partial_z\kappa\ge0$, and then \eqref{eq:cfwad} is a convex combination of values of $\partial_am$. When $Y=g(A,\varepsilon)$ with $\varepsilon\indep(A,Z)\mid T$, $\E\{\partial_ag(A,\varepsilon)\mid A,T\}=\partial_am(A,T)$ and \eqref{eq:cfwad} also equals $\E\{\omega(Z,T)\partial_ag(A,\varepsilon)\}/\E\{\omega(Z,T)\}$, a weighted average of individual structural derivatives. The weight is identified from exposure and instrument data once $T$ is, since $\kappa$ is then the inverse of $T(\cdot,z)$.

\paragraph{Imbens--Newey model.} With $A=h(Z,\eta)$, $\eta\indep Z$ scalar and $h$ strictly increasing in $\eta$, $T=F_{A\mid Z}(A\mid Z)=F_\eta(\eta)$ satisfies $A\indep\varepsilon\mid T$ for any outcome disturbance $\varepsilon$ with $(\varepsilon,\eta)\indep Z$ (their Theorem 1), $T\indep Z$, and $A=\kappa(T,Z)=Q_{A\mid Z}(T\mid Z)=h(Z,F_\eta^{-1}(T))$, so $\partial_z\kappa=\partial_zh(Z,\eta)$, the individual's responsiveness to the instrument, identified as $\partial_zQ_{A\mid Z}(T\mid Z)$.

\paragraph{Existence of an ECoF does not imply that $F_{A\mid Z}(A\mid Z)$ is one: a counterexample.}
Consider the triangular model $Y=g(A,\varepsilon)$, $A=h(Z,\eta)$, $(\varepsilon,\eta)\indep Z$, with the first stage unrestricted, and take
\[
\eta\sim U(0,1),\quad Z\sim U(0,2)\ \text{independent},\qquad A=(Z-1)\,\eta,\qquad Y=m(A,\eta)+e,\quad e\indep(\eta,Z),
\]
so that $\varepsilon=(\eta,e)$, $g(a,\varepsilon)=m(a,\eta)+e$ and $Y(a)=m(a,\eta)+e$.

\emph{$T=\eta$ is an ECoF.} Since $Z=1$ has probability zero, $T=\eta=A/(Z-1)$ is a function of $(A,Z)$; $T\indep Z$ by construction; and $A=\kappa(T,Z)$ with $\kappa(t,z)=(z-1)t$, differentiable in $z$ with $\partial_z\kappa=T\ge0$. For the outcome condition, given $T$ the exposure $A=(Z-1)T$ is a function of $Z$, so $Y(a)\indep(A,Z)\mid T$ reduces to $Y(a)\indep Z\mid T$; and $Y(a)=m(a,T)+e$ with $e\indep(T,Z)$, so $\E\{b(Y(a))\mid T,Z\}=\E\{b(m(a,T)+e)\mid T\}$ for every bounded $b$. Hence \eqref{eq:cf} holds with $T=\eta$, the causal curve is $m(a,t)=\E\{Y(a)\mid T=t\}$, and \eqref{eq:cfwad} applies with $\omega=\partial_z\kappa\,w(Z)/f_Z(Z)=\eta\{1-(Z-1)^2\}/2$ for this uniform instrument, a nonnegative weight.

\emph{$V=F_{A\mid Z}(A\mid Z)$ is not an ECoF.} For $Z=z>1$, $A=(z-1)\eta$ is increasing in $\eta$, so $V=\pr\{(z-1)\eta'\le(z-1)\eta\mid\eta\}=\eta$; for $z<1$ it is decreasing in $\eta$, so $V=\pr\{(z-1)\eta'\le(z-1)\eta\mid\eta\}=\pr(\eta'\ge\eta\mid\eta)=1-\eta$. Thus $V=\eta$ when $Z>1$ and $V=1-\eta$ when $Z<1$: $V$ is uniform and independent of $Z$, as it must be, but it is a function of $(\eta,Z)$, not of $\eta$. Conditional on $V=v$, therefore, $\eta=v$ among those with $Z>1$ and $\eta=1-v$ among those with $Z<1$, and
\[
\E\{Y(a)\mid V=v,\,Z>1\}=m(a,v),\qquad \E\{Y(a)\mid V=v,\,Z<1\}=m(a,1-v),
\]
which differ whenever $m(a,t)\neq m(a,1-t)$. So $Y(a)\indep Z\mid V$ fails and $V$ is not an ECoF, although an ECoF exists. The first stage $h(z,\eta)=(z-1)\eta$ is strictly monotone in $\eta$ for every $z\neq1$, but in a direction that depends on $z$; the proof of Theorem 1 of Imbens and Newey writes $\pr\{h(z,\eta)\le x\}=\pr\{\eta\le h^{-1}(z,x)\}$ and so uses a direction common to all $z$. The same failure arises without any reversal of direction when $\eta$ has disconnected support and $h(z,\cdot)$, while injective, orders the pieces of the support differently at different values of $z$. More generally, $V$ generates the same conditioning as a given ECoF $T$ exactly when $t\mapsto\kappa(t,z)$ orders the support of $T$ in the same way for almost every $z$, which is the Imbens--Newey model. A simulation of the example with $m(a,t)=\sin a+at+t^2$ ($n=6\times10^6$) gave $\beta_{\IV}=1.616$ against $1.618$ from \eqref{eq:cfwad} with $T=\eta$, and $\E\{Y(0.5)\mid V\approx0.2\}$ equal to $0.62$ among $Z>1$ and $1.51$ among $Z<1$. Every ECoF is a $z$-dependent measure-preserving rearrangement $\psi_Z(V)$ of $V=F_{A\mid Z}(A\mid Z)$, since both are injective in $a$ given $z$ and independent of $Z$; which rearrangement, if any, is causal is not identified from the law of $(Y,A,Z)$, any such law being consistent with $V$ or with any other rearrangement satisfying \eqref{eq:cf}. The Imbens--Newey choice is therefore the natural default, and the location shift gives it; the ECoF model states the assumption as the existence of a causal rearrangement rather than as a particular one.

For a normal instrument $w/f_Z$ is constant and \eqref{eq:cfwad} is the responsiveness-weighted average derivative $\E\{\partial_zh\,\partial_am\}/\E\{\partial_zh\}$, the continuous-exposure analogue of compliance weighting~\cite{ImbensAngrist1994,AngristGraddyImbens2000}. Heterogeneous responsiveness, as from gene--environment interaction in the first stage, reweights individuals and is the individual-level form of the variant-specific weighting. A simulation with $A=Z+e^{0.3Z}\eta$, $g(a,\varepsilon)=\sin a+a\varepsilon+\varepsilon^2$ and $\varepsilon$ correlated with $\eta$ gave $\beta_{\IV}=0.531$ against $0.530$ from \eqref{eq:cfwad} for a normal instrument and $0.715$ against $0.716$ for a uniform one, while the plain average derivative was $0.37$ and $0.50$.

\paragraph{Location shift.} With $A=\mu(Z)+V$, $V\indep Z$, take $T=A-\mu(Z)=V$; then $\kappa(t,z)=\mu(z)+t$, $\partial_z\kappa=\mu'(z)$, and $\omega(z)=\mu'(z)w(z)/f_Z(z)$ depends on $Z$ alone, is nonnegative when $\mu$ is nondecreasing, and \eqref{eq:cfwad} reads $\beta_{\IV}=\E\{\omega(Z)h'(A,U)\}/\E\{\omega(Z)\}$ also in terms of the latent confounder $U$, because $\E\{\partial_am(A,T)\mid A,U,T\}$ averages $h'(A,U)$ within $T$ and $\omega$ is a function of $Z$ alone; directly, $\cov(Z,Y)=\cov\{Z,H(\mu(Z))\}$ with $H(m)=\E\{h(m+V,U)\}$ and $H'(m)=\E\{h'(m+V,U)\}$. The model is the Imbens--Newey model with $h(z,\eta)=\mu(z)+\eta$, and $F_{A\mid Z}(A\mid Z)=F_V\{A-\mu(Z)\}$ is a strictly increasing transformation of $A-\mu(Z)$, so the two control functions generate the same conditioning. In the notation of \eqref{S:owndose}, with $S_V(v\mid u)=\pr(V>v\mid U=u)$ and $f_V(\cdot\mid u)$ its density, $g(a,u)=\cov\{Z,S_V(a-\mu(Z)\mid u)\}=\int\omega(z)f_Z(z)f_V\{a-\mu(z)\mid u\}\dd z=f(a\mid u)\,\E\{\omega(Z)\mid A=a,U=u\}$, so $\rho(a,u)=\E\{\omega(Z)\mid A=a,U=u\}$. NUCEM with respect to $U$ fails in general for nonlinear $h$, because $g(a,u)=f(a\mid u)\rho(a,u)$ varies with $u$ whenever the distribution of $V$ given $U$ does, although $\gamma_A(u)=\cov\{\mu(Z),Z\}/\var(Z)$ is constant, so that under \eqref{eq:linear} NOSH's covariance condition, hence NUCEM, holds. For an allele count with $\mu_k=\mu_0+\beta k$, the discrete form of \eqref{eq:cfwad} is
\begin{equation}\label{eq:allele}
\beta_{\IV}=\frac{w_0\theta_0+w_1\theta_1}{w_0+w_1},\qquad \theta_k=\frac{\E\{Y(\mu_{k+1}+V)-Y(\mu_k+V)\}}{\beta}=\frac{H(\mu_{k+1})-H(\mu_k)}{\beta},
\end{equation}
the mean effect per unit dose of shifting everyone's exposure by one allele's worth, averaged over the two allele contrasts.

\paragraph{Proportional perturbation.} If $\omega\equiv c$, \eqref{eq:cfwad} gives $\beta_{\IV}=\E\{\partial_am(A,T)\}$; conversely, if $\E[\{\omega(Z,T)-c\}\partial_am(A,T)]=0$ for every $m$, taking $\partial_am=\omega-c$ gives $\omega\equiv c$ almost surely, so the condition is necessary and sufficient for the plain average derivative to be the estimand for every outcome model. Since $w(z)/f_Z(z)$ is constant exactly for a normal instrument (it equals $\var(Z)$ by Stein's identity), under a normal instrument $\omega$ is constant if and only if $\partial_z\kappa$ is, that is, $\kappa(t,z)=\beta z+\phi(t)$: a location shift with linear $\mu$. In lens-1 terms the condition is $\rho\equiv c$, equivalently $g(a,u)=\cov(Z,A)f(a\mid u)$, and the same argument applied to \eqref{S:owndose} shows it necessary and sufficient for $\beta_{\IV}=\E\{h'(A,U)\}$ for every outcome model. Integrating $g(a,u)=cf(a\mid u)$ over $a$ gives $\cov(A,Z\mid U=u)=c$ for every $u$, so proportional perturbation is equivalent to constant within-stratum instrument strength together with $g(a,u)/\cov(A,Z\mid U=u)=f(a\mid u)$: the instrument's effect on exceeding each dose, normalized by its effect on the mean, coincides with the stratum's exposure density. Equivalently, since $\int_{-\infty}^{a}g(s,u)\dd s=\cov\{Z,\E(\min(A,a)\mid U=u,Z)\}$, it states that $\cov\{Z,\E(\min(A,a)\mid U=u,Z)\}=\cov(Z,A)\,\pr(A\le a\mid U=u)$ for all $a,u$. For normal $Z$ with variance $\sigma^2$, Stein's identity gives $g(a,u)=\sigma^2\E_Z\{\partial_zS(a\mid u,Z)\}$, so the condition reads $\E_Z\{\partial_zS(a\mid u,Z)\}=\beta f(a\mid u)$ with $\beta=c/\sigma^2$; the equal location shift $S(a\mid u,z)=S_V(a-\beta z\mid u)$ satisfies it pointwise, since $\partial_zS=\beta f_V(a-\beta z\mid u)$ averages over $Z$ to $\beta f(a\mid u)$. Under it $\E\{\partial_am(A,T)\}=\E\{h'(A,U)\}$, the mean of each individual's slope at the individual's exposure. NUCEM and proportional perturbation are logically independent: the location shift with a normal instrument satisfies the latter but not the former for nonlinear $h$, and the additive density with a weight $g$ of varying shape satisfies the former but not the latter.

\paragraph{Linear heterogeneous outcome.} If $\E\{Y(a)\mid T=t\}=\beta_0(t)+\beta_Y(t)a$ then $\partial_am(A,T)=\beta_Y(T)$ and, by $T\indep Z$ and Yitzhaki's identity, $\E\{\omega(Z,T)\mid T=t\}=\int\partial_z\kappa(t,z)w(z)\dd z=\cov\{Z,\kappa(t,Z)\}=\cov(A,Z\mid T=t)$, so \eqref{eq:cfwad} becomes \eqref{eq:cflinear}, and $\beta_{\IV}-\E\{\beta_Y(T)\}=\cov_T\{\beta_Y(T),\gamma_A(T)\}/\E\{\gamma_A(T)\}$, which is \eqref{eq:noshgap} with $T$ in the role of $U$: NUCEM with the control function as the confounder is necessary and sufficient for $\beta_{\IV}=\ACE$. Under the location shift $\gamma_A(t)=\cov\{\mu(Z),Z\}/\var(Z)$ is constant and the condition holds automatically; in the Imbens--Newey model it fails whenever $\beta_Y(T)$ covaries with $\E\{\partial_zh(Z,\eta)w(Z)/f_Z(Z)\mid T\}$. A simulation in the Imbens--Newey model above with $\beta_Y(T)=0.5+T$ gave $\beta_{\IV}=1.089$ against $1.089$ from \eqref{eq:cflinear} and $\ACE=1.000$.

\paragraph{Nonlinear additive outcome.} If $\E\{Y(a)\mid T\}=h(a)+\nu(T)$ then $\partial_am(A,T)=h'(A)$ and
\[\E\{\omega(Z,T)h'(A)\}=\int h'(a)\,\E\{\omega(Z,T)\mid A=a\}\,f_A(a)\dd a.\]
Since $T$ satisfies conditions (ii)--(iii), lens 1 applies with $U=T$, and no effect modification gives NUCEM, so $\beta_{\IV}=\int h'g/\int g$ with $g(a)=\cov\{Z,\ind(A>a)\}$; equating the two representations for every $h'$ gives $\E\{\omega(Z,T)\mid A=a\}f_A(a)=g(a)$, so the two lenses agree, with the same identified weight, while linearity fails. A simulation with $h'(a)=\cos a+0.3a$ in the Imbens--Newey model gave $\beta_{\IV}=0.449$, $0.449$ from lens 1 and $0.449$ from lens 2.

\paragraph{NUCEM under the ECoF model.} With $T$ in the role of the confounder, the within-stratum weight is $g(a,t)=\cov\{Z,\ind(\kappa(t,Z)>a)\}$, which varies strongly with $t$ because exposure is a deterministic function of $Z$ within $T=t$; NUCEM with respect to $T$ therefore fails in general when $\partial_am(a,t)$ depends on $t$, and holds when it does not, which is the nonlinear additive case above.

\paragraph{The benign expansion \eqref{eq:benign}.}
From \eqref{eq:allele} with $w_0=2p(1-p)^2$, $w_1=2p^2(1-p)$ and $\theta_k=H'(\mu_0)+H''(\mu_0)\beta(k+\tfrac12)+O(\beta^2)$, and $w_1/(w_0+w_1)=p$, we get $\beta_{\IV}=H'(\mu_0)+H''(\mu_0)\beta(p+\tfrac12)+O(\beta^2)$. The population mean of the genetic component is $\bar m=\mu_0+2p\beta$, so $H'(\mu_0)=H'(\bar m)-2p\beta H''(\bar m)+O(\beta^2)$, and $\beta_{\IV}=H'(\bar m)+H''(\bar m)\beta(\tfrac12-p)+O(\beta^2)$. Finally $H'(\bar m)=\E\{h'(\bar m+V,U)\}=\E\{h'(A,U)\}+O(\beta^2)$ since $A-\bar m-V=\beta(Z-2p)$ has mean zero. For $h$ quadratic in $a$ the expansion is exact. A simulation with $h(a,u)=0.4(a-1)^2+0.3ua$, $V=0.8+0.6U+0.8\varepsilon$ and
$(\beta,p)$ equal to $(0.1,0.3)$, $(0.3,0.1)$, $(0.5,0.45)$ and $(0.8,0.05)$ reproduced \eqref{eq:allele} and the deviations $H''\beta(\tfrac12-p)=0.016,0.096,0.020,0.288$ to Monte Carlo accuracy, and for a normal instrument returned $\beta_{\IV}=-0.159$ against $\E\{h'(A,U)\}=-0.159$.

\section{Pipeline estimands under nonlinearity}
\label{app:pipeline}
\setcounter{equation}{0}

Let $n_Y$ be the outcome sample size, so that $\mathrm{se}(\hat\gamma_{Y,j})^2\approx\var(Y)/\{n_Y\var(G_j)\}$, and let $\omega_j\propto\gamma_{A,j}^2/\mathrm{se}(\hat\gamma_{Y,j})^2\propto\gamma_{A,j}^2\var(G_j)$ be the fixed-effect inverse-variance weights, normalized to sum to one. Then
\[
\theta_{\mathrm{IVW}}=\sum_j\omega_j\theta_j=\int h'(a)\bar g(a)\dd a,\qquad \bar g(a)=\sum_j\omega_j\frac{g_j(a)}{\int g_j},
\]
a weighted average slope with a weight that is a precision-weighted mixture of the variant weights. Cochran's $Q=\sum_j\omega_j(\hat\theta_j-\hat\theta_{\mathrm{IVW}})^2$ has noncentrality
\[
\sum_j\frac{(\theta_j-\theta_{\mathrm{IVW}})^2}{\mathrm{se}(\hat\theta_j)^2},\qquad
\mathrm{se}(\hat\theta_j)^2\approx\frac{\var(Y)}{n_Y\gamma_{A,j}^2\var(G_j)},
\]
which grows linearly in $n_Y$. In the benign case, $\theta_j-\theta_{\mathrm{IVW}}\approx H''\{\beta_j(\tfrac12-p_j)-\mathrm{const}\}$; with $H''=0.8$, $\beta_j=0.05$, $p_j=0.3$ and $n_Y=500{,}000$, the deviation is $0.008$ against $\mathrm{se}(\hat\theta_j)\approx0.045$, so each variant contributes $(0.008/0.045)^2\approx0.03$ to $Q-\mathrm{df}$; with effect sizes spread between $0.02$ and $0.1$, allele frequencies between $0.05$ and $0.5$ and $n_Y$ in the millions, $Q/\mathrm{df}$ of 1.2--1.8 ($I^2$ of 15--45\%) arises from curvature alone. MR-Egger regresses $\hat\gamma_{Y,j}$ on $\hat\gamma_{A,j}$ with intercept; in the benign case, with $\bar\theta=\E\{h'(A,U)\}$, $\gamma_{Y,j}\approx\bar\theta\gamma_{A,j}+H''(\tfrac12-p_j)\gamma_{A,j}^2$, so with effect alleles oriented positive the fitted intercept is nonzero and the slope depends on the distribution of $(\gamma_{A,j},p_j)$; in general $(\theta_j-\bar\theta)\gamma_{A,j}$ is proportional to $\gamma_{A,j}$, violating InSIDE. The weighted median converges to the $\omega$-weighted median of $\{\theta_j\}$ and the mode to the modal cluster of $\{\theta_j\}$.

\section{Observational equivalence of summary statistics}
\label{app:equiv}
\setcounter{equation}{0}

\noindent\emph{Claim.} Any collection of summary statistics $\{(\gamma_{A,j},\gamma_{Y,j})\}$ with $\gamma_{A,j}\neq0$, with any allele frequencies, is generated exactly both by (i) valid instruments satisfying the additive-density condition, with nonnegative weights and a nonlinear dose-response, and by (ii) a linear constant-effect dose-response with a direct effect of each variant on the outcome.

(ii) Let $A=\sum_j\gamma_{A,j}G_j+U+\eta$ and $Y=\theta A+\sum_j\alpha_jG_j+U+\varepsilon$ with independent variants and errors. Then $\cov(G_j,A)/\var(G_j)=\gamma_{A,j}$ and $\cov(G_j,Y)/\var(G_j)=\theta\gamma_{A,j}+\alpha_j$, which equals $\gamma_{Y,j}$ on taking $\alpha_j=\gamma_{Y,j}-\theta\gamma_{A,j}$, for any $\theta$.

(i) Let $\theta_j=\gamma_{Y,j}/\gamma_{A,j}$ and choose disjoint intervals $I_1,\dots,I_L$. Let a selector $R$ take the value $j$ with probability $\pi_j$, independent of $(G,U)$, the value $u$ with probability $\pi_u(U)$, and $0$ otherwise. Given $R=j$, draw $A$ from a location family supported in $I_j$ with location $\lambda_j+\delta_jG_j$; given $R=u$ from $f_u(\cdot\mid U)$; given $R=0$ from $f_0$. The conditional density $\sum_j\pi_jf_j(a\mid g_j)+\pi_u(u)f_u(a\mid u)+\{1-\sum_j\pi_j-\pi_u(u)\}f_0(a)$ satisfies the additive-density condition in each $g_j$. Writing $S_j(a\mid g)$ for the survival function of the route-$j$ density at allele count $g$, each $g_j(a)=\pi_j\cov\{G_j,S_j(a\mid G_j)\}$ is nonnegative, supported in $I_j$, and integrates to $\cov(G_j,A)=\pi_j\delta_j\var(G_j)$, so $\gamma_{A,j}$ is matched by $\pi_j\delta_j=\gamma_{A,j}$. Let $Y=h(A)+U+\varepsilon$ with $h$ smooth and $h'=\theta_j$ on $I_j$; by representation \eqref{eq:wad}, which holds under the additive density, $\cov(G_j,Y)/\cov(G_j,A)=\theta_j$, so $\gamma_{Y,j}$ is matched. If the $\theta_j$ are not all equal, $h$ is nonlinear.

\section{Hybrid design}
\label{app:hybrid}
\setcounter{equation}{0}

For $\phi_k$ with antiderivative $\Phi_k$, $\int\phi_k(a)g_j(a)\dd a=\cov\{G_j,\int\phi_k(a)\ind(A>a)\dd a\}=\cov\{G_j,\Phi_k(A)\}$ up to a constant that does not contribute to the covariance. If NUCEM holds for each $G_j$ with the same population slope $h'=\sum_kc_k\phi_k$, representation \eqref{eq:wad} gives $\cov(G_j,Y)=\int h'g_j=\sum_kc_k\cov\{G_j,\Phi_k(A)\}$, which is \eqref{eq:hybrid} after writing $\cov(G_j,Y)=\gamma_{Y,j}\var(G_j)$ with $\var(G_j)=2p_j(1-p_j)$. Identification of $c$ requires the $L\times K$ matrix $M_{jk}=\cov\{G_j,\Phi_k(A)\}$ to have rank $K$. Under the location shift with small $\beta_j$, $\cov\{G_j,\Phi_k(\beta_jG_j+V)\}=\beta_j\var(G_j)\E\{\phi_k(A)\}+O(\beta_j^2)$ for every $k$, so $M$ has rank one to first order and only $\sum_kc_k\E\{\phi_k(A)\}=\E\{h'(A)\}$ is identified. In the illustration of Figure~1 with the J-shaped curve, the $3\times2$ system for $\phi=(1,a)$ built from variants 1--3 returns $(c_0,c_1)=(-0.65,0.80)$, the true values, with zero residual. With $L>K$ and estimated quantities, the weighted residual sum of squares of \eqref{eq:hybrid} is asymptotically $\chi^2_{L-K}$ under the joint hypothesis.

\section{Illustration parameters}
\label{app:illus}
\setcounter{equation}{0}

$U\in\{0,1\}$ with equal probability. Exposure is drawn, independently of everything else given the route, from: variant $j$'s route with probability $\pi_j$ ($0.15,0.10,0.10,0.10$), a confounder route with probability $0.2+0.1U$ from $N(1.5U-0.5,0.8^2)$, and otherwise from $N(0.8,1)$. Route densities given allele count $g$: variant 1, $N(0.8+0.25g,1)$; variant 2, $N(-0.8+0.4g,0.4^2)$; variant 3, $N(2.6+0.4g,0.4^2)$; variant 4, $N(0.8+0.1g,(0.5+0.35g)^2)$. Allele frequencies $0.3,0.2,0.2,0.4$. Dose-responses: J-shaped, $h'(a,u)=0.8(a-1)+0.3u$; sigmoid, $h'(a,u)=(1+0.2u)\,2\varphi\{(a-1.6)/0.7\}/0.7$ with $\varphi$ the standard normal density. All quantities in Table~2 are computed by numerical integration; standard errors in Figure~1 use $\mathrm{se}(\hat\gamma_{Y,j})^2=1/\{n_Y\var(G_j)\}$ with $n_Y=300{,}000$ and treat $\gamma_{A,j}$ as known. Code reproducing the illustration and all numerical checks accompanies the manuscript.

\section{When are the weights nonnegative? Comonotonicity and non-crossing distribution functions}
\label{app:mono}
\setcounter{equation}{0}

\paragraph{Lens 1 (Result 3(i)).} The weight $g(a)=\cov\{Z,\ind(A>a)\}=\cov\{Z,S(a\mid Z)\}$ is nonnegative for every $a$ whenever $S(a\mid z)=\pr(A>a\mid Z=z)$ is nondecreasing in $z$ for every $a$: $Z$ and $S(a\mid Z)$ are then comonotone functions of $Z$, and the covariance of comonotone functions is nonnegative (for independent copies $Z,Z'$, $2\cov\{Z,S(a\mid Z)\}=\E[(Z-Z')\{S(a\mid Z)-S(a\mid Z')\}]\ge0$). Conversely, if $S(a_0\mid z_1)>S(a_0\mid z_2)$ for some $z_1<z_2$, an instrument distribution concentrated on $\{z_1,z_2\}$ with probabilities $\pi_1,\pi_2$ gives $g(a_0)=\pi_1\pi_2(z_2-z_1)\{S(a_0\mid z_2)-S(a_0\mid z_1)\}<0$, so monotonicity is necessary for nonnegativity to hold under every instrument distribution. The within-stratum form, $S(a\mid u,z)$ nondecreasing in $z$ for every $(a,u)$, gives $g(a,u)\ge0$ and makes the no-NUCEM representation \eqref{S:owndose} a convex combination of individual-level slopes; it implies the marginal form, since $S(a\mid z)=\E\{S(a\mid U,z)\}$. The condition says that higher values of the variant shift the exposure distribution upward in the sense of first-order stochastic dominance: the conditional distribution functions $F(\cdot\mid z)=1-S(\cdot\mid z)$ do not cross.

\paragraph{Lens 2 (Result 3(ii)).} Since $w\ge0$, $\omega(z,t)\ge0$ if and only if $\partial_z\kappa(t,z)\ge0$. In the Imbens--Newey model $\kappa(t,z)=Q_{A\mid Z}(t\mid z)=\inf\{a:F(a\mid z)\ge t\}$. If $F(a\mid z)$ is nonincreasing in $z$ for every $a$, then for $z_1<z_2$ the set $\{a:F(a\mid z_2)\ge t\}$ is contained in $\{a:F(a\mid z_1)\ge t\}$, so $Q(t\mid z_2)\ge Q(t\mid z_1)$ for every $t$; conversely, if $z\mapsto Q(t\mid z)$ is nondecreasing for every $t$, then $F(a\mid z)=\sup\{t:Q(t\mid z)\le a\}$ is nonincreasing in $z$ for every $a$. Hence $\partial_z\kappa\ge0$ for every $t$ if and only if $S(a\mid z)$ is nondecreasing in $z$ for every $a$: the two lenses share one condition, non-crossing of the conditional distribution functions of exposure given the variant. Under the location shift, $\partial_z\kappa=\mu'(z)$ and the condition is that $\mu$ be nondecreasing. For the location--scale first stage $A=\mu(Z)+\sigma(Z)\eta$ with $\eta\indep Z$, $\kappa(t,z)=\mu(z)+\sigma(z)F_\eta^{-1}(t)$ and $\partial_z\kappa=\mu'(z)+\sigma'(z)F_\eta^{-1}(t)$, which is negative for $t$ small enough whenever $\sigma'(z)>0$ and $\eta$ is unbounded below. The distribution functions at $z_1<z_2$ then cross at the dose $a^\ast$ solving $\{a^\ast-\mu(z_1)\}/\sigma(z_1)=\{a^\ast-\mu(z_2)\}/\sigma(z_2)$: for $a<a^\ast$, $S(a\mid z_2)<S(a\mid z_1)$, so $g(a)<0$, and for $a>a^\ast$, $g(a)>0$. This is the structure of variant 4 in Figure~1, whose ratio exceeds every individual slope for the sigmoid dose-response.

Checkable primitives. Write $F(a\mid z)=1-S(a\mid z)$ and $f(a\mid z)$ for the conditional distribution and density functions of $A$ given $Z=z$, and $F_0$, $f_0$ for a fixed baseline distribution and density. Then $S(a\mid z)$ is nondecreasing in $z$ for every $a$ under any of the following: (i) a location shift, $F(a\mid z)=F_0\{a-\eta(z)\}$ with $\eta$ nondecreasing, which includes the additive first stage of setting (c); (ii) a monotone likelihood ratio, $\partial^2_{az}\log f(a\mid z)\ge0$, which implies first-order dominance; (iii) a transformation model $T(A)=\eta(Z)+\xi$ with $T$ strictly increasing, $\eta$ nondecreasing and $\xi\indep Z$, since dominance is invariant to increasing transformations; (iv) a location--scale model $F(a\mid z)=F_0[\{a-\mu(z)\}/\sigma(z)]$ with $\mu'(z)+\sigma'(z)v\ge0$ for all $v$ in the support of the standardized error, which follows from $\partial_zF(a\mid z)=-f_0(v)\{\mu'(z)+\sigma'(z)v\}/\sigma(z)$ at $v=\{a-\mu(z)\}/\sigma(z)$. Under (iv) the variant may affect spread as well as location provided the location effect dominates on the support; with an unbounded standardized error only a pure location shift preserves monotonicity at every dose, and a sufficiently strong scale effect produces crossing distributions and a weight of both signs.

By the equivalence above, non-crossing holds if and only if every conditional quantile $z\mapsto F^{-1}(\tau\mid z)$, $\tau\in(0,1)$, is nondecreasing, which is how it can be checked with individual-level genotype and exposure data: fit quantile regressions of exposure on allele count across a grid of $\tau$ and test whether all slopes are nonnegative, or, with a three-valued allele count, compare the empirical survival functions of exposure at the three genotypes and look for crossings. A variant that passes has a nonnegative weight, and its ratio, under NUCEM, is a proper average of the dose-response slope over the doses where its weight is concentrated; a variant that fails has no such guarantee under any assumption on the outcome. Within-stratum monotonicity, the condition of setting (d), can be probed in the same way within observed covariate strata, though not within strata of the unmeasured $U$.

\end{document}